\documentstyle[prd,aps,epsf,graphicx,psfrag]{revtex}

\newcommand{\half}{\mbox{$\textstyle \frac{1}{2}$}}

\newcommand{\beq}{\begin{equation} }
\newcommand{\eeq}{\end{equation} }
\newcommand{\beqa}{\begin{eqnarray} }
\newcommand{\eeqa}{\end{eqnarray} }

\begin{document}

\tighten
\draft
\preprint{}
\twocolumn[\hsize\textwidth\columnwidth\hsize\csname
@twocolumnfalse\endcsname

\title{Brane Gas Inflation}

\author{
M. F. Parry$^{*}$ and D. A. Steer$^{\dagger}$
}

\address{* Theoretical Physics Group, Blackett Laboratory, Imperial
College, London SW7 2BZ, UK}

\address{$\dagger$ D\'epartement de Physique Th\'eorique, Universit\'e de 
Gen\`eve,
24 Quai Ernest Ansermet, 1211 Gen\`eve 4, Switzerland\\
and\\
Laboratoire de Physique Th\'eorique, B\^at. 210, Universit\'e Paris XI,
91405 Orsay Cedex, France}

\date{September 21, 2001}

\maketitle

\begin{abstract}
We consider the brane gas picture of the early universe. At later stages,
when there are no winding modes and the background is free to expand, we
show that a moving 3-brane, which we identify with our universe, can
inflate even though it is radiation-dominated. The crucial ingredients for
successful inflation are the coupling to the dilaton and the equation of
state of the bulk. If we suppose the brane initially forms in a collision 
of higher-dimensional branes,
then the spectrum of primordial density fluctuations naturally has a
thermal origin.
\end{abstract}

\pacs{PACS Numbers : 98.80.Cq, 11.27.+d \hspace{2cm}
Imperial/TP/0-01/37
\qquad ORSAY-LPT-01-86}

\vskip2pc]

In recent work, Alexander {\it et al.} \cite{abe} proposed a dynamical
origin of the non-compact spatial dimensions of the universe. In their
picture, the universe starts as a hot, dense gas of the fundamental states
of string theory, namely D-branes. Considering 11-dimensional M-theory
compactified on $S^1$, they show that winding modes will allow only some
spatial dimensions to grow large; a result which generalizes that of
Brandenberger and Vafa\cite{bv}. The relevant equations of motion (see also
\cite{tv}) emphasize the importance of the dilaton to this proposal.

Here we consider what might be the late-time behavior of such a universe.
We will suppose that $d$ spatial dimensions have become large and that all
degrees of freedom that have been able to interact have annihilated with
one another. In particular, we assume that there are no more winding modes
so that the universe is free to expand. In a similar set-up, Park {\it et
al.}\cite{psl} envisaged the universe as a gas of D$p$-branes in the
context of Brans-Dicke theory. As for our model, we will be more precise
shortly as to the dominant contribution to the energy density of the
universe, but for now we imagine that are a number of 3-branes in this
universe. Furthermore, we will suppose that what we think of as {\em our}
universe is, in fact, one of these branes. Our proposal, therefore, is
similar in spirit to that of mirage cosmology\cite{kk}.

In this paper, we show that a 3-brane, moving in this background, can
inflate. This is true even though the brane, assumed formed in a
collision, is radiation-dominated.  Consequently, primordial density
fluctuations are seen to be thermal in origin. The parameter space of
inflationary solutions is spanned by the coupling to the dilaton and the
bulk barotropic index. The set-up has elements in common with those of
Alexander\cite{steff} and Burgess {\it et al.}\cite{ddbar}. However, in
our case inflation is not due to brane---anti-brane interaction and we
require the 3-brane to be moving rather quickly.

In our scenario, the time evolution of our universe is governed both by
the matter on the brane and its dynamics in the expanding background.
However, we do not consider the self-gravity of the brane, this being
still an open problem. Solving for the brane dynamics then is analogous
to determining planetary motion in that we assume a background and do not
take back-reaction into account. In particular, we do not have the usual
brane world junction conditions which are, in any case, difficult to
apply to objects with codimension greater than one. We start by
considering the background.

If we assume the background is flat and roughly homogeneous and isotropic
in the
$d$ spatial dimensions, then the metric can be written as
\beq
g_{\mu\nu}  = {\mbox {diag}}\{1,-a^2(t),\ldots,-a^2(t)\},
\eeq
where $\mu = 0 \ldots d$ and we will let $x^{\mu}$ label the coordinates,
with $t \equiv x^0$. Following \cite{abe}, 
we take the low-energy effective action of the bulk to be the
dilaton-gravity action in
$D=d+1$ dimensions:
\beq
S_B = \int d^D x \sqrt{-g} \left\{ \frac{e^{-2\phi}}{2 \kappa_D^2}\left[
R +
4(\nabla\phi)^2 \right] + {\cal L}_B \right\},
\eeq
where $\phi$ is the dilaton, ${\cal L}_B$ describes the matter in the
bulk, $\kappa_D$ is related to the $D$-dimensional Newton's
constant in the usual way and we have supposed there is no bulk
cosmological constant.

Taking the bulk matter to be of perfect fluid form, the
equations of motion for the background are
\beqa\label{bulk1}
\half d (d\!-\!1) H^2 &=& e^{2\phi} \kappa_D^2\,\rho - 2\dot{\phi}
(\dot{\phi}\!-\!d H) \\ \label{bulk2}
(d\!-\!1) \dot{H} + \half d (d\!-\!1) H^2 &=& -e^{2\phi} \kappa_D^2\,p+
2(\ddot{\phi}\!-\!H\dot{\phi}) \nonumber \\
{}&&\hspace{0.5cm}-
2\dot{\phi}(\dot{\phi}\!-\!d H)\\
\label{bulk3}
d \dot{H} + \half d (d\!+\!1) H^2 &=&
2\ddot{\phi}-2\dot{\phi}(\dot{\phi}\!-\!d
H), 
\eeqa
where $\rho$ is the density and $p$ the pressure of the fluid, and
$H=\dot{a}/a$. The case
of pure Einstein gravity is recovered if we let $\phi=0$ and drop
(\ref{bulk3}).

As usual, the conservation equation, here $\dot{\rho} + d H
(\rho\!+\!p) = 0$, follows from the field equations or it can be
derived from $\nabla_{\nu}T^{\mu\nu}=0$. Thus, if we augment the system
(\ref{bulk1})-(\ref{bulk3}) with an equation of state $p=w\rho$, we
find that
\beq
\rho \sim a^{-d(1+w)}.
\eeq

This suggests the following ansatz for the dilaton which is indeed the
general solution at low curvature scales\cite{gv}:
\beq
e^{2\phi} \sim a^n,
\eeq
where $n$ is a constant. The equations of motion are
satisfied only if
\beq\label{soln}
n=d -\frac{1}{w}\qquad {\mbox{and}} \qquad a \sim
t^{\frac{2}{d(1+w)-n}} = t^{\frac{2w}{1+dw^2}}.
\eeq
Thus, the background expands for all $w>0$ but it is not hard to
see that bulk
inflation is not possible for any choice of $w$ and $d$. This is not in
contradiction with models of dilaton-driven inflation because we have
assumed the background has evolved into a low curvature regime. It
should also be pointed out that the case of $w=0$ is not pathological.
Even though $n\!\rightarrow\! -\infty$, the density and scale factor
simply become constant while $e^{2\phi} \sim t^{-2}$. However,
we will have no use for $w<0$ solutions since, as we will see, the bulk
and the brane expand or contract together. Having said that, $w<0$ is
allowed in pure Einstein gravity; there $n=0$ and $w$ and $d$ are
unconstrained. Bulk inflation occurs for $|w+1| < 2/d$.

Having determined the evolution of the bulk, we now turn to the
question of the action for the brane. We are led by the usual
Dirac-Born-Infeld action but will make some
simplifications. We will suppose that the brane is not charged
under any fields living in the bulk and, as in \cite{abe}, we will assume
that usual antisymmetric tensor field $B_{\mu\nu}$
vanishes. Additionally, we adopt a
phenomenological approach to enable us to put ordinary forms of matter on
the brane. We suppose the brane action is:
\beq
S_b =  \int d^4\sigma \sqrt{-\gamma} {\cal L} =
\int
d^4\sigma \sqrt{-\gamma} \{ e^{-\phi}\lambda + \xi e^{-m\phi}
{\cal L}_b\},
\eeq
where $\sigma^i$ are the world-volume coordinates of the brane ($i=0
\ldots 3$), $\gamma_{ij}$ the induced metric, $\lambda$ the tension of
the brane, and $m$ and $\xi$ are dimensionless constants which determine
the coupling of the
dilaton to the brane matter given by ${\cal L}_b$. Note that this action
allows us to consider the usual gauge fields that might live on the brane
(by expanding the usual square-root term in powers of the string
tension) but that it does not include the effects of the brane
self-gravity.

The induced metric on the 3-brane follows from its embedding in the
bulk, i.e. $x^{\mu}\!=\!X^{\mu}(\sigma)$. We choose the
static gauge $\sigma^i\!=\!x^i$ and suppose an embedding of the form
\beq
X^i = x^i, \qquad X^A = X^A(t),
\eeq
where $A=4 \ldots d$. Then the induced metric on the brane is
\beqa
\gamma_{ij} &\equiv& g_{\mu\nu} \frac{\partial X^{\mu}}{\partial \sigma^i}
\frac{\partial X^{\nu}}{\partial \sigma^j} \nonumber \\
&=& {\mbox{diag}}\{1-V^2,-a^2,-a^2,-a^2\},
\eeqa
where $V^2 \equiv -\dot{X}^A\dot{X}_A \geq 0$. Note that the scale
factor on
the brane is the same as that in the bulk. However, because of the
motion of the brane in the transverse directions, the brane time
$\tau$, defined via
\beq\label{time}
d\tau = dt \sqrt{1-V^2},
\eeq
is not the same as the bulk time. Accordingly, a brane-bound observer will
see a quite different evolution of the scale factor, with the discrepancy
becoming more pronounced as $V^2\!\rightarrow\!1$. We are
especially interested in 
the possibility of having the brane inflate even though the bulk does not. 
Therefore, we now consider the brane dynamics.

Since ${\cal L}_b$, for consistency, cannot contain spatial
derivatives and nor does it depend on $X^A$ (only $\dot{X}^A$ through
$\gamma_{00}$), it follows that $\partial \sqrt{-\gamma}{\cal L}/\partial
\dot{X}^A$ are constants of the brane motion. To be precise,
\beq
c_A=\frac{\partial \sqrt{-\gamma}{\cal L}}{\partial \dot{X}^A} =
\frac{\partial \gamma_{00}}{\partial \dot{X}^A} \frac{\delta
\sqrt{-\gamma}{\cal L}}{\delta \gamma_{00}} = \sqrt{-\gamma} \dot{X}_A
T^{00}
\eeq
are constant. It follows that
\beq\label{const}
c \equiv \sqrt{-c^A c_A} = \frac{a^3 V}{\sqrt{1-V^2}} T^0{}_0
\eeq
is a positive constant, where
\beq
T^0{}_0 = e^{-\phi} \lambda + \xi e^{-m\phi} \rho_b
\eeq
and $\rho_b$ is the energy density of the matter on the
brane. Equation (\ref{const}) can be inverted to find $V$ which is then
inserted into
(\ref{time}). We find
\beq\label{time2}
d\tau = \frac{a^3 T^0{}_0}{\sqrt{c^2 + a^6 (T^0{}_0)^2}}\, dt.
\eeq

Before applying this equation to a definite scenario, a small mathematical
digression will be helpful. It is not hard to show that if
\beq
d\tau \sim a^{\beta} \, dt \qquad {\mbox{and}} \qquad a \sim
t^{\frac{1}{\alpha}},
\eeq
then
\beq
a \sim (\epsilon \tau)^{\frac{1}{\alpha+\beta}} \qquad {\mbox{where}}
\qquad \epsilon = {\mbox{sgn}}(\alpha\!+\!\beta).
\eeq
Thus, we will have brane inflation when $|\alpha\!+\!\beta| <
1$ and, in particular, exponential inflation when $\alpha\!+\!\beta=0$. We
will use the variables $\alpha$ and $\beta$ in what follows.

We are now in a position to consider a particular physical situation. Our
idea is that the 3-brane which is our universe came about as a result of a
collision process, say a 5-$\bar{5}$ brane annihilation\cite{steff,ms}. We
make two
assumptions about this collision. Firstly, that the resulting velocity of
the 3-brane in the transverse directions is, at least initially, 
relativistic. In other words
\beq\label{ass1}
c \gg a^3 T^0{}_0.
\eeq
Furthermore, we suppose a large amount of energy is deposited on the
brane in the collision and that this dominates over the brane tension,
i.e. 
\beq\label{ass2}
\rho_b \gg \lambda
\eeq
or, to be slightly more accurate, $\xi e^{-m\phi}
\rho_b \gg e^{-\phi}\lambda$.

The equation of state for brane matter most consistent with a
collision event is $p_b\!=\!\rho_b/3$ which corresponds to radiation on
the brane or the
excitation of massless scalar modes.
If, as is usual, we assume the matter on the brane remains
confined to it, then energy conservation determines that
\beq\label{ass3}
\rho_b \sim a^{-4}.
\eeq
From (\ref{soln}) and substituting (\ref{ass1})-(\ref{ass3}) into
(\ref{time2}), we have that
\beq\label{exp}
\alpha +\beta = \half[d (1+w) - n(1+m)] -1.
\eeq

The parameter space of solutions is quite constrained. On M-theoretic
grounds we expect $d\!\leq\!10$ and we need $d\!\geq\!4$ in order for the
background to be able even to contain a moving 3-brane. However, if our
universe is the result of 5-$\bar{5}$ brane annihilation, then we would
require $d\!>\!5$. Furthermore, if there are a number of 3-branes in the
bulk, then we would prefer our universe to avoid them. Since two
$p$-branes
will interact in at most $2p\!+\!1$ large spatial dimensions, it follows
that we must have $d\!>\!7$, $d\!=\!7$ being marginal.

Next, we will limit ourselves to $0\!<\!w\!\leq\!1$. The lower bound is
required to have expansion at all and the upper bound is the usual case
of stiff matter. Actually, we might expect $w$ to be no greater than $1/d$
which would correspond to radiation in the bulk. In the case of Einstein
gravity, we will consider $-1\!\leq\!w\!\leq\!1$.

We now search for inflationary solutions. The case of minimal coupling of
the dilaton to brane matter can be dealt with immediately. Putting
$m\!=\!0$ into (\ref{exp}), we find $|\alpha\!+\!\beta|<1$ when $dw^2
\!-4w\!+\!1 < 0$, but this only has solutions for $d\!<\!4$. Thus,
inflation in this scenario will require non-minimal coupling to the
dilaton.

\begin{figure}[thb]
\psfrag{y}[tb]{\hspace{-0.5cm}$m$}
\psfrag{x}[bt]{$w$}
\psfrag{0}[c]{$0$}
\psfrag{0.1}[c]{}
\psfrag{0.2}[c]{$0.2$}
\psfrag{0.3}[c]{}
\psfrag{0.4}[c]{$0.4$}
\psfrag{0.5}[c]{}
\psfrag{0.6}[c]{$0.6$}
\psfrag{0.7}[c]{}
\psfrag{0.8}[c]{$0.8$}
\psfrag{0.9}[c]{}
\psfrag{-1}[c]{$-1$}
\psfrag{-2}[c]{$-2$}
\psfrag{-3}[c]{}
\psfrag{1}[c]{$1$}
\psfrag{2}[c]{$2$}
\psfrag{3}[c]{$3$}
\includegraphics[width=8cm,angle=0]{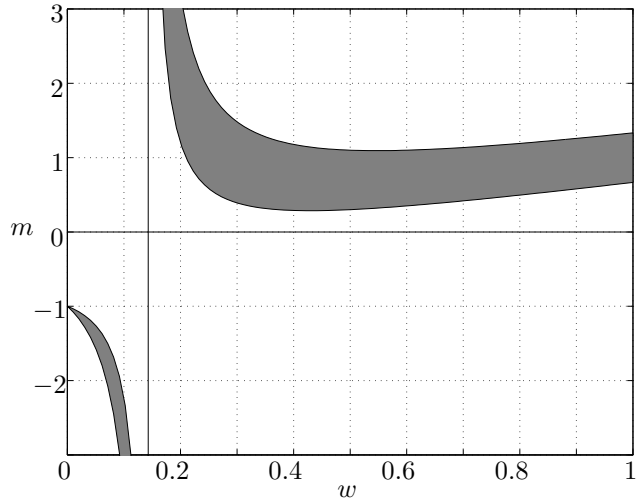}
\vspace{0.5cm}
\caption{The shaded regions indicate inflationary scenarios for a 3-brane
moving in $d\!=\!7$ spatial dimensions. The regions grow as $d$ is
decreased, and shrink as $d$ is made larger. Parameter space is
spanned
by $m$,
which gives the coupling of the dilaton to brane matter,
and $w$, the
ratio of the bulk pressure to the bulk density.}
\label{fig1}
\end{figure}

The typical effect of the coupling to the dilaton is illustrated in fig.
\ref{fig1}. The number of spatial dimensions has been set to seven and the
shaded regions indicate the values of $(w,m)$ which give rise to inflation
on the brane. These regions grow as $d$ gets smaller, and shrink for
larger values of $d$. Their behavior as $w\!\rightarrow\!1/d$, in other
words $n\!\rightarrow\!0$, is due to the dilaton becoming independent
of the scale factor. The interesting result here is that it is
possible to obtain brane inflation with ordinary matter both on the
brane and in the bulk.

A natural question to ask is how inflation ends in this scenario. It
turns out that in cases of successful inflation (\ref{ass1}) remains
valid but the brane tension eventually becomes dominant, reversing
(\ref{ass2}). Now we have $\alpha\!+\!\beta=d(1+w)/2\!-\!n\!+\!3$ and a
simple analysis reveals it is not possible to have inflation for the range
of parameters we are considering. However, this fact allows us to make a
straightforward estimation of the number of $e$-folds of inflation,
$N$. If we suppose inflation ends when $e^{-\phi}\lambda = \xi e^{-m\phi}
\rho_b$, then
\beq\label{efolds}
N = \frac{\ln{\xi} + \ln{(\frac{\rho_0}{\lambda})} -
(m-1)\phi_0}{\half 
n(m\!-\!1)+4},
\eeq
where subscript $0$ indicates the value of the quantity at the start of
inflation and we are now letting $\rho$ be the density of matter on
the brane. The
denominator is positive for inflationary solutions but it diverges to
infinity as $w\!\rightarrow\!0$. Since we can expect realistically 
$\xi\!\sim\!1$ and 
$\ln{(\rho_0/\lambda)}\!\lesssim\!10$, we see that the
initial
value of the dilaton will, in general, be crucial to ensuring the
usual requirement for inflation that $N\!\gtrsim\!70$.

For example, when $w\!>\!1/d$, $n$ is positive so that the value of the
dilaton when the 3-brane is formed can be arbitrarily large. Furthermore,
as seen in fig. \ref{fig1}, we can have inflation for $m\!<\!1$, thus
the large initial value of the dilaton translates to a large value of $N$.
On the other hand, an interesting case arises when $w\!\lesssim\!1/d$.
Then the size of $\phi_0$ can be irrelevant because $m$ must large and
negative for inflation to occur. However, we expect such values of $m$ to
be theoretically unlikely.

The fact that we do not have inflation when $\lambda$ dominates may seem
a little odd until we remember we have neglected the self-gravity of the
brane in our analysis\footnote{Indeed, if one allows $w_b$ to vary as
well, the absence of self-gravity means inflation becomes more likely as
$w_b\!\rightarrow\!1$, and not, as one might expect, as $w_b$ becomes more
negative. It is even possible to have inflation without any coupling to
the dilaton in this case, with inflation ending because (\ref{ass1})
becomes no longer satisfied.}. Without doubt, until this is included,
details of the all-important transition from inflation to radiation
domination will not be known and we will be unable to accurately determine
the number of $e$-folds of inflation. However, although it is still an
open problem how to address the issue of self-gravity, it seems plausible
that at early brane times the assumptions (\ref{ass1}) and (\ref{ass2})
mean that the motion of the brane through its background is more
important than its internal dynamics in determining the evolution of the
brane. It is tempting, however, to suppose that when self-gravity effects
become important, it may be that $\lambda$, attenuated by its coupling to
the dilaton, generates late-time acceleration on the brane.

The construction of inflationary solutions is quite different in the case
of pure Einstein gravity. It follows when $n\!=\!0$ is substituted into
(\ref{time2}) that brane inflation can only occur for $w$ negative.
However, there is an interesting region $-1+2/d < w < -1+4/d$ where the
brane inflates but the bulk does not. Once again, inflation ends when the
brane tension starts to dominate the
matter density on the brane. In this instance though, it will be difficult
to have sufficient $e$-folds of inflation. This can be seen by putting
$n=\phi_0=0$ in (\ref{efolds}); we require $\rho_0$ to be bigger than
$\lambda$ to a fantastic degree in order to have $N$ large. Therefore, it
appears that dilaton gravity is essential to the inflationary scenarios we
have considered here.

One of the more compelling features of the usual models of inflation is
that they naturally give rise to primordial density fluctuations. In the
scenario presented here this is also the case. There are two
contributions. The first are thermal fluctuations on the brane, coming
from the collision, for which the spectrum will likely need to be quite
red in order to fit with observations. The details will depend crucially on
the manner of the formation of the 3-brane\cite{ps}. The second
contribution comes from bulk fluctuations which are then induced on the
brane. The need here will be for an account of the background evolution.
And of course, in both cases, self-gravity will have to be included in
order to trace the primordial spectrum through to the present day.

To summarize, we have proposed that our universe could be a 3-brane moving
in a late-time, brane gas background. Of particular note is that this
brane can, depending on the details of the coupling of brane matter to the
dilaton and the nature of the bulk matter, successfully inflate.
Furthermore, if the brane is a result of a collision, there appears to be
a natural mechanism by which density fluctuations would arise on the
brane.

There are a number of avenues worth exploring however. Firstly, one could
include explicitly the gauge fields which live on the brane. Since the
usual interpretation is that these fields reflect the existence of open
strings ending on the brane, one is further led to consider brane-brane
interactions. Perhaps the ``near miss'' of another 3-brane during the
evolution of our universe is responsible for the varying of the fine
structure constant? Secondly, one might suppose the 3-brane is charged
under bulk Ramond-Ramond fields. On a more phenomenological level, one
could consider different couplings to the dilaton and a varying bulk
equation of state. Indeed, it is not to hard to see that brane inflation
could begin or end because of a change in $w$. Lastly, the most pressing
need is to understand the effect of brane self-gravity. We have argued
that this may not be important at early brane times, however it will be
vital if one is to turn our proposal into a fully viable cosmological
model. One requirement will be that the self-gravity is confined to the
brane so as not to violate Newton's Law.

It is a pleasure to thank Raul Abramo for helpful discussions. MP
gratefully acknowledges the hospitality of the theoretical physics group
at Ludwig-Maximilians Universit\"{a}t where this work was completed.

$*$ Email: {\tt parry@theorie.physik.uni-muenchen.de }

$\dagger$ Email: {\tt steer@th.u-psud.fr }

\end{document}